\documentclass[reprint,aps,showpacs]{revtex4-1}
\usepackage{dcolumn}
\usepackage[dvipdfm]{hyperref}
\usepackage[dvipdfm]{graphicx}
\usepackage{amsmath}
\usepackage{hypernat}
\begin{document}
\title{Room-temperature high-speed nuclear-spin quantum memory in diamond}

\author{J. H. Shim}
\author{I. Niemeyer}
\author{J. Zhang}
\author{D. Suter}
\affiliation{
 Fakult\"{a}t Physik, Technische Universit\"{a}t Dortmund, D-44221 Dortmund, Germany\\
}

\begin{abstract}

Quantum memories provide intermediate storage of quantum information until it is needed for the next step of a quantum
algorithm or a quantum communication process.
Relevant figures of merit are therefore the fidelity with which the information can be written and retrieved,
the storage time, and also the speed of the read-write process.
Here, we present experimental data on a quantum memory consisting of a single $^{13}$C nuclear spin that is strongly
coupled to the electron spin of a nitrogen-vacancy (NV) center in diamond.
The strong hyperfine interaction of the nearest-neighbor carbon results in  transfer times of 300 ns between the register qubit
and the memory qubit, with an overall fidelity of 88 \% for the write - storage - read cycle.
The observed storage times of 3.3 ms appear to be limited by the $T_1$ relaxation of the electron spin.
We discuss a possible scheme that may extend the storage time beyond this limit.
\end{abstract}


\maketitle

\section{Introduction}

Storage for quantum information is one of the essential parts of quantum computing and
quantum communication \cite{NC01, Stolze:2008xy}, and the essential features of quantum memories
have been demonstrated in different physical implementations \cite{Julsgaard2004, Rabl2006, Morton2008, Lvovsky2009, Timoney2011, Fuchs2011,Maurer2012}.
Within these different implementations, the nitrogen-vacancy (NV) center in diamond has the particular appeal of room temperature operation\cite{Jelezko2004,Dutt2007,Neumann2008}
and long coherence times \cite{Balasubramanian2009}. These properties make it a promising candidate for quantum information processing,
an interface for quantum communication, and also as a nanoscale sensor for electric and magnetic fields \cite{Maze2008a, Dolde2011}.
While most of these applications rely on the electron spin of the NV center as the qubit,
the center also contains nuclear spins, whose extended coherence times
have advantages for applications such as quantum memories \cite{Fuchs2011,Maurer2012}.
Such nuclear spin quantum memories can be  integrated into quantum computing architectures that use NV centers in diamond \cite{Barrett2005, Stoneham2009}.

If nuclear spins are used as quantum memories, another relevant parameter is the coupling strength between
the processing qubit (the electron spin) and the memory qubit (the nuclear spin).
The strength of this interaction determines the duration of the information transfer between the
two qubits; stronger couplings generally result in faster operation speed and therefore a lower loss of fidelity
from decoherence.
We can broadly distinguish between strongly coupled systems, where the coupling strengths $A$ are larger
than the Rabi frequencies  $\Omega_R$, and weakly coupled systems, where $A < \Omega_R$.
While earlier experiments have explored weakly coupled systems, it is the purpose of this paper to discuss
a strongly coupled system.
In addition to the advantage of short gate times, strongly coupled systems can provide simpler gate operations.
As an example, the controlled-NOT (CNOT) operation, which is the central part of the information transfer between
the two qubits, can be implemented by a single control pulse in a strongly coupled system.

Figure~\ref{rabi} (a) shows the strongly coupled quantum memory that we want to explore in the present study.
It consists of a diamond NV center coupled to a  $^{13}$C nuclear spin  in the nearest neighbor site.
The hyperfine interaction between these spins is $\approx 130$ MHz \cite{Jelezko2004, Neumann2008, Smeltzer2011},
which is much larger than our typical Rabi frequencies of $25$ MHz.

In this paper, we first discuss the system and introduce a purification scheme of the two-qubit system
by a cycle of laser and microwave pulses, which significantly increases the purity of the initial state.
We then demonstrate that the strong coupling also enhances the operation speed of the single-qubit operations
on the nuclear spin by more than two orders of magnitude.
On this basis, we explore the storage of quantum information in the memory qubit, determine the fidelity of the read-write process
and protect the stored information against environmental noise by dynamical decoupling.

\section{System and  Setup}

\begin{figure}
\includegraphics[width=\columnwidth]{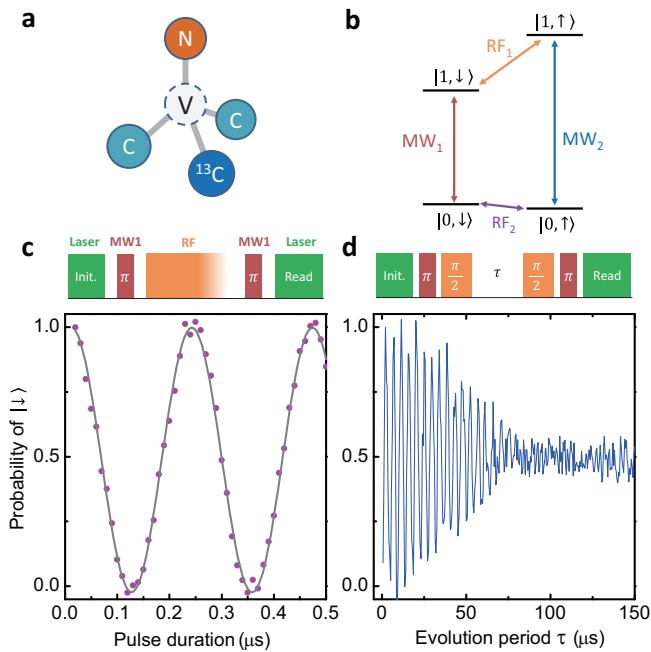}
\caption{(Color online)
(a) Structure of the NV center with one $^{13}$C nuclear spin in the first shell.
(b) Energy levels and transitions of the subsystem spanned by the $m_S=0$ and  $m_S=1$
states of the electron spin and the two $^{13}$C nuclear spin states.
(c) Pulse sequence for measuring the  $^{13}$C Rabi oscillation and the resulting signal.
(d) Pulse sequence for measuring the $^{13}$C FID and the resulting signal}
\label{rabi}
\end{figure}

We consider the quantum register of  Fig.~\ref{rabi} (a), which consists of a single electron spin ($S=1$)  and
a single $^{13}$C nuclear spin ($I=1/2$).
For the present work, the $^{14}$N nuclear spin is not relevant and will not be considered.
The relevant Hamiltonian is
\begin{equation*}
\mathcal{H} = D S_z ^2 + \gamma_e B S_z + A_{\|}S_z I_z + A_{\perp}(S_x I_x + S_y I_y).
\end{equation*}
Here, the $S_{\alpha}$ are the electron spin operators,  $I_{\alpha}$ the nuclear spin operators,
$D$ is the zero field splitting $ \gamma_e B$ the strength of the Zeeman interaction and
$A_{\|}$ and $A_{\perp}$ the secular and nonsecular parts of the hyperfine coupling.
We label the eigenstates of this bipartite system as $|m_S,\uparrow\rangle$ and $|m_S,\downarrow \rangle$,
where $|\uparrow\rangle$ and $|\downarrow\rangle$ represent the nuclear spin  states.
Figure~\ref{rabi} (b) shows the energy level scheme and the relevant transitions.
Since the hyperfine coupling $A_{\|}$ is significantly stronger than the Rabi frequency,
pulses applied to the transitions  MW$_1$, MW$_2$, and RF$_1$, are to an excellent approximation,
selective for the targeted transitions.
The resonance frequency for microwave and rf transitions RF$_1$ is 127.2 MHz and RF$_2$ was not used in this work.

For the experiments described here, we used a type IIb natural abundance diamond crystal,
whose nitrogen impurity concentration is $<5$ ppb, and a home-built confocal microscope.
A diode-pumped solid-state laser with an emission wavelength of 532 nm induced the laser field
for the optical excitation.
The cw laser beam was sent through an acousto-optical modulator with an extinction ratio of 58 dB and a
rise time of 50 ns to generate the laser pulses for excitation and detection.
The microwave pulses for the excitation of the electron spin were generated from
two direct digital synthesis (DDS) sources, passed through a switch with an isolation of 90 dB
and a 16 W amplifier.
The radio-frequency pulses for driving the nuclear spin were generated with another DDS source
and passed through a switch and a 50 W amplifier.
Microwave and rf excitation pulses were combined with a power combiner and passed through a Cu wire
mounted on the surface of the diamond crystal.
For these experiments, we aligned the field of a permanent magnet (65 G) to the symmetry axis of the NV center,
with the precision $<0.1^{\circ}$.

\section{Experimental Results}

\subsection{Ultrafast manipulation of the $^{13}$C nuclear spin}

An important precondition for storage and readout is the control of the qubit by external fields.
We therefore determined the strength of the interaction with the resonant radio-frequency (rf)
field by measuring Rabi oscillations, using the pulse sequence shown in Fig.1 (c).
The first laser pulse (duration $> 10 \mu$s) initializes the electron spin into the $m_S=0$ (bright) state.
Under our experimental conditions (small magnetic field), it also  fully depolarizes the nuclear spin
(see Sec. \ref{sec_init.} for details).
The subsequent $\pi$ pulse on the MW$_1$ transition populates the $|1,\downarrow\rangle $ state
and the rf pulse drives the $|1,\downarrow\rangle \leftrightarrow |1,\uparrow\rangle $ transition,
whose Rabi frequency is what we want to measure.
The second microwave pulse brings the remaining population of $|1,\downarrow\rangle$ back into the
$m_S=0$ (bright) state, where it can be read out with the second laser pulse.

Figure~\ref{rabi} (c)  shows the result of this measurement.
The vertical axis represents the probability that the system remains in $|1,\downarrow\rangle$.
The value of 0 indicates that the spin has been completely transferred to  $|1,\uparrow\rangle$.
The fitted cosine curve in the figure corresponds to a Rabi frequency of 4.3 MHz.
Driving a bare $^{13}$C nuclear spin at this rate would require an rf field strength of 428.5 mT,
which is more than two orders of magnitude higher than the value of the experimentally applied field.

This strong enhancement of the transition dipole moment is a consequence
of the strong hyperfine coupling that does not commute with the electron spin Hamiltonian.
As a result, the eigenstates of the combined system are not the product states of the subsystems.
While this mixing is small (of the order of 0.03), the transition dipole moment also receives a
corresponding admixture of the electron spin dipole moment.
These enhanced dipole moments have been used recently in the context of hybrid quantum registers
\cite{Morley2012}. A similar case of spin - orbit qubits induced by a strong spin-orbit coupling was reported
in rare-earth ions \cite{Bertaina2009} and other examples are known in NMR of magnetic materials \cite{Turov}.

According to the authors of Ref. \cite{Maze2008}, we expect an enhancement by a factor
$\frac{\gamma_e}{\gamma_N} \frac{A_{\perp}}{D}(3 |m_S|-2)$,
where $\gamma_e$  and  $\gamma_N$ are the  gyromagnetic ratios of electron and nuclear spin, respectively.
Using $A_{\perp} \approx 127$ MHz \cite{Shim_Prepare}, we expect an enhancement factor of $\approx 125$,
and a numerical simulation of the nuclear spin Rabi oscillation yields a value of 122.
From the data in the figure, we find that the duration of an rf $\pi$ pulse is $\approx 125$ ns.
This pulse, which is selective with respect to the electron spin state, corresponds to a C$_e$NOT$_n$ operation required for transferring  coherence between the electron and nuclear spin qubits.
This gate duration is much shorter than the coherence time ($T_2 ^*$) of the NV electron spin of $\approx 1 \mu$s.
We therefore expect that  transferring the quantum state between the two qubits should be possible
with no significant loss of fidelity.

\begin{figure}[t]
\includegraphics[width=\columnwidth]{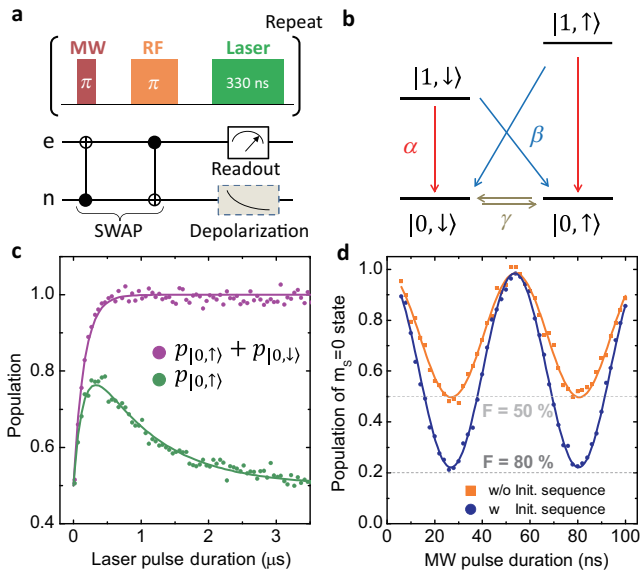}
\caption{(Color online) \textbf{Initialization scheme and  resulting fidelity}
(a) Pulse sequence used.
Since the states $|1,\downarrow\rangle$ and  $|1,\uparrow\rangle$ are initially both unpopulated,
one MW$_1$ and one RF$_1$ $\pi$ pulse are sufficient to generate a swap operation.
The laser  pulse
polarizes the electron spin and partly depolarizes the nuclear spin.
(b) Rate equation model for analyzing the effect of the laser illumination on the populations.
(c) Experimental  (points) and calculated  (curves) populations as a function of the laser pulse duration.
(d) Rabi oscillations with and without  ten cycles of the initialization sequence.
With (without) the initialization sequence, the fidelity is 80 \% (50 \%).}
\label{initialization}
\end{figure}

\subsection{Initialization}\label{sec_init.}

In the ideal quantum storage experiment, we should start with the two-qubit system in the $|0,\downarrow\rangle $ state.
The usual initialization by a laser pulse leads to a large polarization of the electron spin,
but to an almost complete depolarization of the nuclear spin.
The population of the desired initial state is therefore only $50$ \%.
Before the actual experiment, we therefore discuss a procedure to optimize the initialization (i.e. a technique for purifying the two-qubit quantum state).

Several possible ways for polarizing the  $^{13}$C  nuclear spins have been discussed before.
One approach consists of applying a magnetic field that brings the excited state
close to the level-anticrossing point \cite{Jacques2009}.
Another approach is the single-shot readout experiment \cite{Neumann2010, Robledo2011, Maurer2012},
but neither of these approaches is feasible under our experimental conditions.
We therefore decided on another approach: We first initialized the electron spin, swaped the polarization
to the nuclear spin and re-initialized the electron spin with a short laser pulse that is long enough to generate
significant polarization of the electron spin, but short enough to retain most of the nuclear spin polarization \cite{Jiang2009}.
This process can then be iterated for optimal polarization of the two-qubit quantum register.
Figure \ref{initialization} (a) shows the corresponding pulse sequence.

\begin{figure}[t]
\includegraphics[width=\columnwidth]{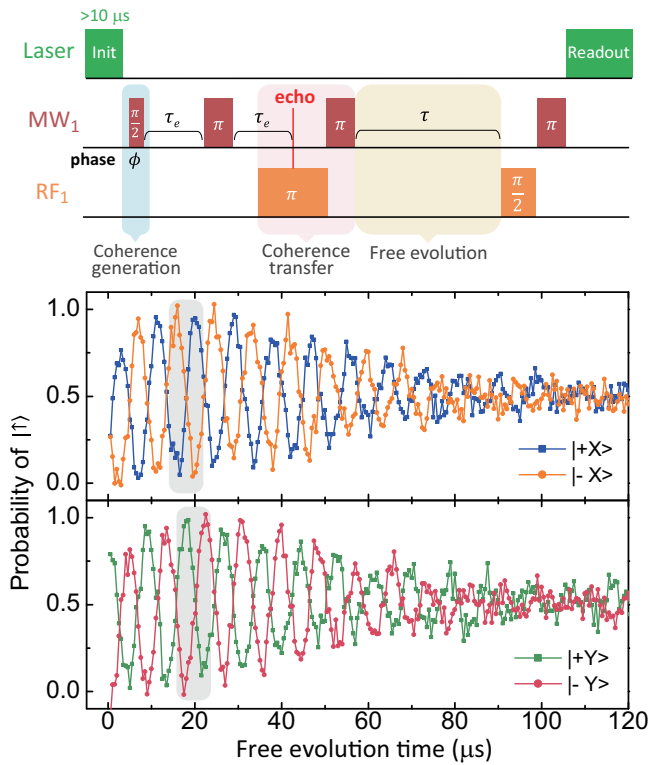}
\caption{(color online) \textbf{Coherence transfer from NV electron spin to $^{13}$C nuclear spin}
Upper part: Pulse sequence for coherence transfer and subsequent free evolution of $^{13}$C.
The phase $\phi$ of the initial $\pi/2$ pulse determines the phase of the superposition state,
which is then transferred to the $^{13}$C quantum memory.
Lower part: Free evolution of $^{13}$C.
The initial phases of the Ramsey fringes reflects the phase $\phi$ of the initial $\pi/2$ microwave pulse.}
\label{transfer}
\end{figure}

To optimize the conditions for this initialization procedure, we first determined the relevant system parameters
for the pumping process during the pulse.
For this purpose, we first initialized the system into a state with a completely polarized nuclear spin,
$\Psi_n(0) = |\uparrow\rangle$, and a completely depolarized electron spin by applying a long
532 nm laser pulse, a microwave $\pi$ pulse to the MW$_1$ transition and an rf $\pi$ pulse to the RF$_1$ transition.
After this initialization, we applied a second laser pulse of variable duration and measured its effect on the populations
of the system by performing a partial population tomography: We read out the total population of the $m_S = 0$ levels
and, in a second experiment, we swapped the $|0, \downarrow\rangle \leftrightarrow |1, \downarrow\rangle$ states
and read out the population of the $|0, \uparrow\rangle$ state.
Figure~\ref{initialization} (c) shows the populations measured in these experiments as a function of the duration
of the laser pulse.
For short durations of the laser pulse, we observe an increase of the population of \emph{both} $m_S = 0$ states.
After $\approx 330$ ns, the population of $| 0,\uparrow \rangle$ reaches a maximum of $\approx 0.78$ and for
long pulse durations, both ground state populations approach the equilibrium value of 0.5.

To analyze the result, we describe the dynamics between the four states under the laser illumination
by the  rate equation model of Fig.~\ref{initialization} (b).
Here, $\alpha$ is the rate for nuclear spin conserving transitions, $\beta$ for nuclear spin flipping transition,
and $\gamma$ for the ground state depolarization.
Using the analytical solution (see the Appendix), we can determine the rate constants by fitting the experimental curves.
We obtained $1/\alpha= 0.17\,\mu$s, $1/\beta=0.92$ $\mu$s, and $1/\gamma =1.6$ $\mu$s.
A higher initial state polarization can be achieved by repeating this initialization cycle several times.
In the experiment, we implemented ten cycles, thereby increasing the initial state population from 50 \% to 80 \%.
This value was also determined by measurements of the Rabi oscillation.
As shown in Fig.~\ref{initialization} (d), the oscillation amplitude rises from 50 $\%$ without the initialization sequence
to $\approx 80$ \% with the repetitive initialization sequence.
For the nuclear spin alone, this initial state corresponds to 60$\%$ polarization.

\begin{figure}[b]
\includegraphics[width=\columnwidth]{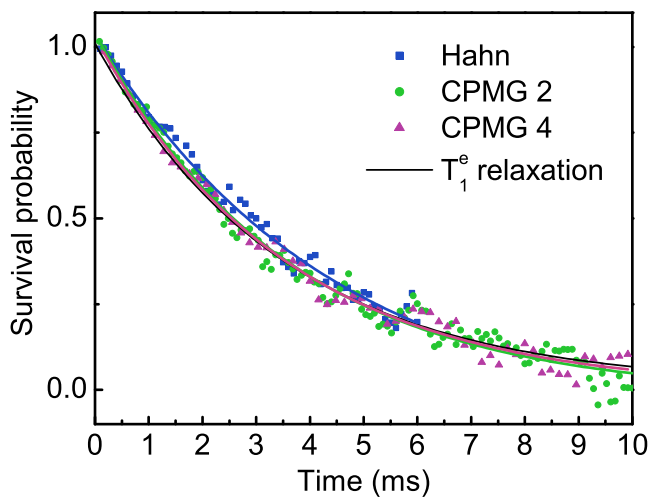}
\caption{(color online)  Survival of $^{13}$C nuclear spin coherence in spin-echo sequences.
The experimental data points are compared to the electron spin-lattice relaxation (black curve).
}
\label{storage}
\end{figure}

\subsection{Storing  coherent superposition states}

We now turn to the actual quantum storage experiment.
For this purpose, we first prepared a coherent superposition state in the electron spin qubit,
transferred this state to the nuclear spin qubit, let it evolve there, and read out the resulting state
through the electron spin.
The upper part of Fig.~\ref{transfer} shows the pulse sequence used for this experiment.
After the initialization sequence, the first $\pi/2$ MW pulse  generates the initial state
$$
|\Psi\rangle (0) = \frac{1}{\sqrt{2}}(|0\rangle + e^{i \phi} |1\rangle).
$$
For ideal pulses, this state could be transferred to the nuclear spin by a simple $\pi$ pulse on the RF$_1$ transition.
Although this pulse is relatively short for a nuclear spin transition, it still leads to a finite loss of fidelity
through relaxation during the transfer and due to the hyperfine interaction with the $^{14}$N nuclear spin.
We further reduced this loss by combining the transfer with a spin-echo sequence, as shown in Fig.~\ref{transfer}.
The timing of the rf $\pi$ pulse was adjusted to align its  center with the echo maximum.
In addition, the gap between the rf $\pi$ pulse and the following mw $\pi$ pulse was kept close to zero.
Those two steps are crucial for high-fidelity  coherence transfer.

To quantify the storage fidelity, the four different initial states,
$|+X\rangle$, $|-X\rangle$, $|+Y\rangle$, and $|-Y\rangle$, corresponding to $\phi = 0, \pi, \pi/2$, and $3\pi/2$
were initialized and stored.
For each state, we measured the free evolution (Ramsey fringes) of the $^{13}$C nuclear spin.
Compared to the reference Ramsey fringe curve in Fig. 1 (d), all  four curves show the same oscillation frequency
and decoherence rates, but  different starting phases.
These starting phases are in accordance with the phases $\phi$ of the initial states prepared
in the electron spin qubit within experimental uncertainty.
This implies that the prepared coherence of the electron spin was successfully transferred to and stored in the nuclear spin.
From the peak-to-peak amplitude ($\Delta$) of the Ramsey fringes, we can determine the fidelity
$\mathcal{F} = (1+\Delta)/2$.\cite{Fuchs2011}
We determined the experimental values for $\Delta$ by fitting a single oscillation in the shaded areas near 20 $\mu$s\footnote{The initial part of the FID is distorted by baseline drifts and compressed amplitudes}.
For the four initial states
$|+X\rangle$, $|-X\rangle$, $|+X\rangle$, and $|-Y\rangle$, we found fidelities of 0.90, 0.83, 0.86, and 0.92, respectively.
The average of these four values is $\overline{\mathcal{F}} = 0.88$.

\begin{figure}[b]
\includegraphics[width=\columnwidth]{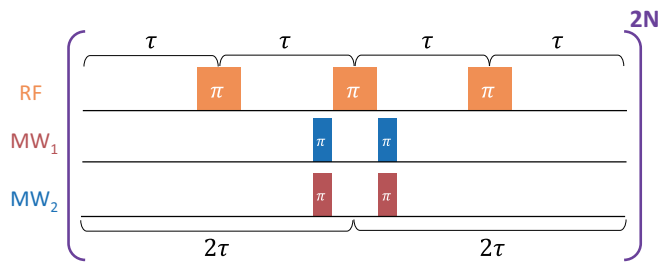}
\caption{(color online) Pulse sequence for a dynamical decoupling scheme that could extend the storage time
beyond the $T_1^\mathrm{e}$ limit.
}
\label{scheme}
\end{figure}

\subsection{Extending the storage time}

According to Fig.~\ref{transfer}, the storage time of the nuclear spin qubit is $\approx 50\,\mu$s.
To further extend this storage time, we applied refocusing pulses to the nuclear spin.
Fig. \ref{storage} (a) shows the amplitude of the echoes generated with 1 (blue dots), 2 (green), and
four (purple) refocusing pulses.
In all cases, the coherence shows an exponential decay with a time constant of $\approx 3.3$ ms.
The fact that the decay time does not depend on the number of refocusing pulses indicates that the
environmental noise that causes the decay of the Ramsey fringes is static on the timescale of these
experiments.
The red solid line compares these decays to the $T_1$ relaxation of the dark state ($m_S=1$),
which is the state of the electron during these storage experiments.
Within experimental uncertainly, the decays are identical, which indicates that the decay is caused
mostly be the relaxation of the electron spin.
The observed lifetime of this quantum memory is significantly longer than the maximally observed lifetimes
of the electron spin coherence in the NV center.
Using dynamical decoupling, the longest reported decoherence times were between 2.2 and 2.5 ms
\cite{Ryan2010, Naydenov2011, Bar-Gill2012, Shim2012}.

\section{Discussion}

\subsection{Decoherence mechanism}

These experimental results demonstrate that quantum mechanical superposition states can be stored for significantly longer times
in the nuclear spin than in the electron spin.
The observed decays appear to be limited by the longitudinal relaxation of the electron spin, rather than by
irreversible dephasing of the nuclear spin.
The observed decay times of the data shown in Fig.\,\ref{storage} are 4.1 $\pm$ 0.52 ms for the Hahn echo,
3.7 $\pm$ 0.31 ms and 3.5 $\pm$ 0.35 ms for the
CPMG-2 and CPMG-4 data.
These values are very close to the value of $T_1^{e} = $ 3.3 $\pm$ 0.2 ms,
which was measured independently.
We expect that the observed decay rate
$$
\frac{1}{T_2} =  \frac{1}{T_2^{C}} + \frac{1}{T_1^{e}}
$$
 is the sum of two independent processes, the pure dephasing of the nuclear spin, with time constant $T_2^{C}$
 and the population decay of the electron spin with time constant $T_1^{e}$.
Since the observed decay time constants are slightly longer than $T_1^{e}$, the pure dephasing time
$T_2^{C}$ must be at least an order of magnitude longer than $T_1^{e}$,
$T_2^{C} > 33$  ms.

\subsection{Storage beyond the T$_ 1^e$ limit}

In the experiment above, we prepared the coherence between the $|1,\uparrow\rangle$ and $|1,\downarrow\rangle$ states
and applied the refocusing pulses to this transition.
The limitation to the coherence lifetime comes from the relaxation of the electron spin:
If the electron spin flips, the nuclear spin coherence is transferred to the
 $|0,\uparrow\rangle \leftrightarrow |0,\downarrow\rangle$ transition, where it precesses with a much lower frequency
 and therefore rapidly looses the phase coherence.
 To avoid this decoherence mechanism, we must also eliminate the dynamical phase acquired in the
 $|0,\uparrow\rangle  \leftrightarrow  |0,\downarrow\rangle$ transition.
 This is possible, in principle, by applying DD pulses to this transition.
 However, the corresponding transition frequency is inconveniently low.
 A significantly more efficient way of dynamical decoupling is therefore to generate the refocusing pulses
 by using the RF$_1$ transition together with the two microwave transitions.

Figure~\ref{scheme} shows the corresponding pulse sequence.
The pairs of $\pi$ pulses on the microwave transitions corresponds to an exchange of the two
nuclear spin transitions.
Together with the RF$_1$ pulse between them, they generate an effective inversion pulse for the
RF$_2$ coherence, which therefore refocuses over the total cycle time of duration $4 \tau$.
The coherence in the RF$_1$ transition is refocused directly, with the first and third RF$_1$ $\pi$ pulses.

\section{Conclusion}

Many tasks in quantum information processing and communication require the intermediate storage of information
in some quantum memory device.
Good quantum memories should therefore excel in a number of aspects, such as long storage, high fidelity,
but also fast read/write times.
Here, we investigated a room-temperature quantum memory consisting of a single $^{13}$C nuclear spin
in the first coordination shell of a NV center in diamond.
This system profits from the strong hyperfine interaction ($\sim$130 MHz) in two ways:
The read-write times are not limited by the strength of the coupling between the two qubits,
but only by the available Rabi frequencies.
Furthermore, the strong hyperfine coupling enhances the nuclear spin Rabi frequency by more than two
orders of magnitude, which also leads to a corresponding speed-up of the gate operations.
In our implementation, the read-write times are limited by the duration of the nuclear spin inversion pulse,
which was 125 ns in our system.
Since this is only slightly shorter than the electron-spin dephasing time, we combined it with a Hahn-echo sequence
on the electron spin, which resulted in a transfer fidelity of 88 \% and a total gate read-write time of 300 ns.
The observed storage time of 3.3 ms appears to be limited by the $T_1$ relaxation of the electron spin.
As we discussed in the text, an extended dynamical decoupling technique should be able to extend the
memory time beyond this limit.

In addition to the actual storage-preservation-recall experiments, we also developed a scheme for optimizing
the polarization of the combined two-qubit system of the electron and nuclear spin, which consists of several cycles
of laser- and microwave pulses.

\appendix

\section{Rate equation model for the effect of laser illumination}

As discussed in the main text, we found that the population dynamics during laser irradiation
can be well described by the rate equation model depicted in Fig. \ref{initialization} (b).
The corresponding equation of motion can be written as
\begin{equation}
\frac{d}{dt}{\vec{P}}(t)= \mathcal{M} \vec{P}(t),
\end{equation}
where the population vector $\vec{P}$ and transition matrix $\mathcal{M}$ are defined as,
\begin{equation}
\vec{P}(t) \hspace{-1mm}=\hspace{-1mm}\begin{pmatrix}P_{|0,\uparrow\rangle} \\ P_{|0,\downarrow\rangle} \\ P_{|1,\uparrow\rangle} \\ P_{|1,\downarrow\rangle}\end{pmatrix}\hspace{-1mm},
\mathcal{M} \hspace{-1mm}=\hspace{-1mm}\begin{pmatrix} -\gamma & \gamma & \alpha & \beta\\
 \gamma & -\gamma & \beta & \alpha \\ 0 & 0 & -(\alpha \hspace{-1mm}+\hspace{-1mm} \beta) & 0 \\0 & 0 & 0 & -(\alpha \hspace{-1mm}+\hspace{-1mm} \beta)\end{pmatrix}
\end{equation}

\begin{figure}[t]
\includegraphics[width=8cm]{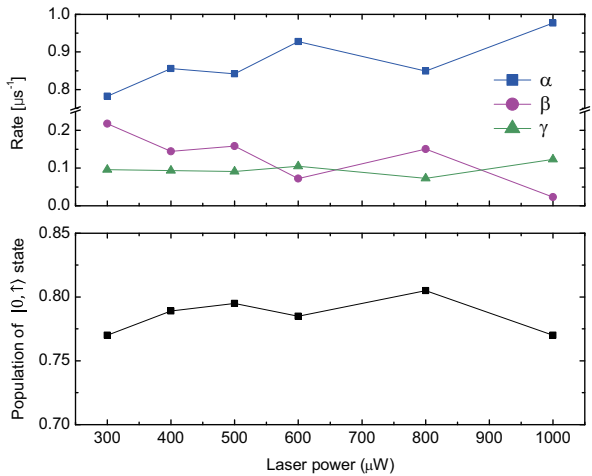}
\caption{(color online)\textit{Upper part} - The measured rates in our rate-equation model of the populations ($\alpha$ (blue), $\beta$ (purple), and $\gamma$ (green))
as a function of laser power.
\textit{Lower part} - Calculated population $P_{|0,\uparrow\rangle}$  after the 4  repetitions of the initialization sequence.}
\label{laser_power}
\end{figure}

The initial conditions, immediately after the first swap operation, are
$P^\top=(0.5,0.5,0,0)$.
With this initial condition, one can easily find the solution for the two populations
$P_{|1,\uparrow\rangle}$ and $P_{|1,\downarrow\rangle}$.
For the remaining populations, we used  the ansatz as $P_{|0,\uparrow\rangle}=\frac{1}{2}+u_{\pm}e^{-(\alpha+\beta)t} \pm \nu e^{-2\gamma t}$  to find the solutions
\begin{equation}
\left\{
\begin{aligned}
P_{|0,\uparrow\rangle}\hspace{-0.5mm}&=\hspace{-0.5mm} \frac{1}{2} \hspace{-0.5mm}-\hspace{-0.5mm}  \frac{\alpha- \gamma}{2(\alpha\hspace{-0.5mm}+\hspace{-1mm}\beta\hspace{-0.5mm}-\hspace{-1mm}2\gamma)}\Big(e^{-(\alpha+\beta)t}\hspace{-1mm}-\hspace{-1mm}e^{-2\gamma t}\Big)\\
P_{|0,\downarrow\rangle}\hspace{-0.5mm}&=\hspace{-0.5mm} \frac{1}{2} \hspace{-0.5mm}-\hspace{-0.5mm}  \frac{\beta-\gamma}{2(\alpha\hspace{-1mm}+\hspace{-1mm}\beta\hspace{-1mm}-\hspace{-1mm}2\gamma)}e^{-(\alpha+\beta)t} \hspace{-0.5mm} -\hspace{-0.5mm} \frac{\alpha-\gamma}{2(\alpha\hspace{-1mm}+\hspace{-1mm}\beta\hspace{-1mm}-\hspace{-1mm}2\gamma)}e^{-2\gamma t}\\
P_{|1,\uparrow\rangle}\hspace{-0.5mm}&=\hspace{-0.5mm} \frac{1}{2}e^{-(\alpha+\beta)t} \\
P_{|1,\downarrow\rangle}\hspace{-0.5mm}&=\hspace{-0.5mm} 0
\end{aligned}
\right.
\label{solution}\end{equation}
We also determined  the three parameters, $\alpha$, $\beta$, and $\gamma$ as a function of the laser intensity.
Figure ~\ref{laser_power} shows the resulting rates in the upper part and the effect of the intensity on the initialization fidelity
in the lower part.
Apparently, the effect is relatively small.

\acknowledgments
This work was supported by the Deutsche Forschungsgesellschaft through Grant No. Su 192/27-1  (FOR 1482).


\begin{thebibliography}{32}
\expandafter\ifx\csname natexlab\endcsname\relax\def\natexlab#1{#1}\fi
\expandafter\ifx\csname bibnamefont\endcsname\relax
  \def\bibnamefont#1{#1}\fi
\expandafter\ifx\csname bibfnamefont\endcsname\relax
  \def\bibfnamefont#1{#1}\fi
\expandafter\ifx\csname citenamefont\endcsname\relax
  \def\citenamefont#1{#1}\fi
\expandafter\ifx\csname url\endcsname\relax
  \def\url#1{\texttt{#1}}\fi
\expandafter\ifx\csname urlprefix\endcsname\relax\def\urlprefix{URL }\fi
\providecommand{\bibinfo}[2]{#2}
\providecommand{\eprint}[2][]{\url{#2}}

\bibitem[{\citenamefont{Nielsen and Chuang}(2001)}]{NC01}
\bibinfo{author}{\bibfnamefont{M.~A.} \bibnamefont{Nielsen}} \bibnamefont{and}
  \bibinfo{author}{\bibfnamefont{I.~L.} \bibnamefont{Chuang}},
  \emph{\bibinfo{title}{Quantum computation and quantum information}}
  (\bibinfo{publisher}{Cambridge Univ. Press, Cambridge},
  \bibinfo{year}{2001}).

\bibitem[{\citenamefont{Stolze and Suter}(2008)}]{Stolze:2008xy}
\bibinfo{author}{\bibfnamefont{J.}~\bibnamefont{Stolze}} \bibnamefont{and}
  \bibinfo{author}{\bibfnamefont{D.}~\bibnamefont{Suter}},
  \emph{\bibinfo{title}{Quantum Computing: A Short Course from Theory to
  Experiment}} (\bibinfo{publisher}{Wiley-VCH, Berlin}, \bibinfo{year}{2008}).

\bibitem[{\citenamefont{Julsgaard et~al.}(2004)\citenamefont{Julsgaard,
  Sherson, Cirac, Fiurasek, and Polzik}}]{Julsgaard2004}
\bibinfo{author}{\bibfnamefont{B.}~\bibnamefont{Julsgaard}},
  \bibinfo{author}{\bibfnamefont{J.}~\bibnamefont{Sherson}},
  \bibinfo{author}{\bibfnamefont{J.~I.} \bibnamefont{Cirac}},
  \bibinfo{author}{\bibfnamefont{J.}~\bibnamefont{Fiurasek}}, \bibnamefont{and}
  \bibinfo{author}{\bibfnamefont{E.~S.} \bibnamefont{Polzik}},
  \bibinfo{journal}{Nature} \textbf{\bibinfo{volume}{432}},
  \bibinfo{pages}{482} (\bibinfo{year}{2004}),
  \urlprefix\url{http://dx.doi.org/10.1038/nature03064}.

\bibitem[{\citenamefont{Rabl et~al.}(2006)\citenamefont{Rabl, DeMille, Doyle,
  Lukin, Schoelkopf, and Zoller}}]{Rabl2006}
\bibinfo{author}{\bibfnamefont{P.}~\bibnamefont{Rabl}},
  \bibinfo{author}{\bibfnamefont{D.}~\bibnamefont{DeMille}},
  \bibinfo{author}{\bibfnamefont{J.~M.} \bibnamefont{Doyle}},
  \bibinfo{author}{\bibfnamefont{M.~D.} \bibnamefont{Lukin}},
  \bibinfo{author}{\bibfnamefont{R.~J.} \bibnamefont{Schoelkopf}},
  \bibnamefont{and} \bibinfo{author}{\bibfnamefont{P.}~\bibnamefont{Zoller}},
  \bibinfo{journal}{Phys. Rev. Lett.} \textbf{\bibinfo{volume}{97}},
  \bibinfo{pages}{033003} (\bibinfo{year}{2006}),
  \urlprefix\url{http://link.aps.org/doi/10.1103/PhysRevLett.97.033003}.

\bibitem[{\citenamefont{Morton et~al.}(2008)\citenamefont{Morton, Tyryshkin,
  Brown, Shankar, Lovett, Ardavan, Schenkel, Haller, Ager, and
  Lyon}}]{Morton2008}
\bibinfo{author}{\bibfnamefont{J.~J.~L.} \bibnamefont{Morton}},
  \bibinfo{author}{\bibfnamefont{A.~M.} \bibnamefont{Tyryshkin}},
  \bibinfo{author}{\bibfnamefont{R.~M.} \bibnamefont{Brown}},
  \bibinfo{author}{\bibfnamefont{S.}~\bibnamefont{Shankar}},
  \bibinfo{author}{\bibfnamefont{B.~W.} \bibnamefont{Lovett}},
  \bibinfo{author}{\bibfnamefont{A.}~\bibnamefont{Ardavan}},
  \bibinfo{author}{\bibfnamefont{T.}~\bibnamefont{Schenkel}},
  \bibinfo{author}{\bibfnamefont{E.~E.} \bibnamefont{Haller}},
  \bibinfo{author}{\bibfnamefont{J.~W.} \bibnamefont{Ager}}, \bibnamefont{and}
  \bibinfo{author}{\bibfnamefont{S.~A.} \bibnamefont{Lyon}},
  \bibinfo{journal}{Nature} \textbf{\bibinfo{volume}{455}},
  \bibinfo{pages}{1085} (\bibinfo{year}{2008}),
  \urlprefix\url{http://dx.doi.org/10.1038/nature07295}.

\bibitem[{\citenamefont{Lvovsky et~al.}(2009)\citenamefont{Lvovsky, Sanders,
  and Tittel}}]{Lvovsky2009}
\bibinfo{author}{\bibfnamefont{A.~I.} \bibnamefont{Lvovsky}},
  \bibinfo{author}{\bibfnamefont{B.~C.} \bibnamefont{Sanders}},
  \bibnamefont{and} \bibinfo{author}{\bibfnamefont{W.}~\bibnamefont{Tittel}},
  \bibinfo{journal}{Nat Photon} \textbf{\bibinfo{volume}{3}},
  \bibinfo{pages}{706} (\bibinfo{year}{2009}),
  \urlprefix\url{http://dx.doi.org/10.1038/nphoton.2009.231}.

\bibitem[{\citenamefont{Timoney et~al.}(2011)\citenamefont{Timoney, Baumgart,
  Johanning, Varon, Plenio, Retzker, and Wunderlich}}]{Timoney2011}
\bibinfo{author}{\bibfnamefont{N.}~\bibnamefont{Timoney}},
  \bibinfo{author}{\bibfnamefont{I.}~\bibnamefont{Baumgart}},
  \bibinfo{author}{\bibfnamefont{M.}~\bibnamefont{Johanning}},
  \bibinfo{author}{\bibfnamefont{A.~F.} \bibnamefont{Varon}},
  \bibinfo{author}{\bibfnamefont{M.~B.} \bibnamefont{Plenio}},
  \bibinfo{author}{\bibfnamefont{A.}~\bibnamefont{Retzker}}, \bibnamefont{and}
  \bibinfo{author}{\bibfnamefont{C.}~\bibnamefont{Wunderlich}},
  \bibinfo{journal}{Nature} \textbf{\bibinfo{volume}{476}},
  \bibinfo{pages}{185} (\bibinfo{year}{2011}),
  \urlprefix\url{http://dx.doi.org/10.1038/nature10319}.

\bibitem[{\citenamefont{Fuchs et~al.}(2011)\citenamefont{Fuchs, Burkard,
  Klimov, and Awschalom}}]{Fuchs2011}
\bibinfo{author}{\bibfnamefont{G.~D.} \bibnamefont{Fuchs}},
  \bibinfo{author}{\bibfnamefont{G.}~\bibnamefont{Burkard}},
  \bibinfo{author}{\bibfnamefont{P.~V.} \bibnamefont{Klimov}},
  \bibnamefont{and} \bibinfo{author}{\bibfnamefont{D.~D.}
  \bibnamefont{Awschalom}}, \bibinfo{journal}{Nat Phys}
  \textbf{\bibinfo{volume}{7}}, \bibinfo{pages}{789} (\bibinfo{year}{2011}),
  \urlprefix\url{http://dx.doi.org/10.1038/nphys2026}.

\bibitem[{\citenamefont{Maurer et~al.}(2012)\citenamefont{Maurer, Kucsko,
  Latta, Jiang, Yao, Bennett, Pastawski, Hunger, Chisholm, Markham
  et~al.}}]{Maurer2012}
\bibinfo{author}{\bibfnamefont{P.~C.} \bibnamefont{Maurer}},
  \bibinfo{author}{\bibfnamefont{G.}~\bibnamefont{Kucsko}},
  \bibinfo{author}{\bibfnamefont{C.}~\bibnamefont{Latta}},
  \bibinfo{author}{\bibfnamefont{L.}~\bibnamefont{Jiang}},
  \bibinfo{author}{\bibfnamefont{N.~Y.} \bibnamefont{Yao}},
  \bibinfo{author}{\bibfnamefont{S.~D.} \bibnamefont{Bennett}},
  \bibinfo{author}{\bibfnamefont{F.}~\bibnamefont{Pastawski}},
  \bibinfo{author}{\bibfnamefont{D.}~\bibnamefont{Hunger}},
  \bibinfo{author}{\bibfnamefont{N.}~\bibnamefont{Chisholm}},
  \bibinfo{author}{\bibfnamefont{M.}~\bibnamefont{Markham}},
  \bibnamefont{et~al.}, \bibinfo{journal}{Science}
  \textbf{\bibinfo{volume}{336}}, \bibinfo{pages}{1283} (\bibinfo{year}{2012}),
  \urlprefix\url{http://www.sciencemag.org/content/336/6086/1283.abstract}.

\bibitem[{\citenamefont{Jelezko et~al.}(2004)\citenamefont{Jelezko, Gaebel,
  Popa, Domhan, Gruber, and Wrachtrup}}]{Jelezko2004}
\bibinfo{author}{\bibfnamefont{F.}~\bibnamefont{Jelezko}},
  \bibinfo{author}{\bibfnamefont{T.}~\bibnamefont{Gaebel}},
  \bibinfo{author}{\bibfnamefont{I.}~\bibnamefont{Popa}},
  \bibinfo{author}{\bibfnamefont{M.}~\bibnamefont{Domhan}},
  \bibinfo{author}{\bibfnamefont{A.}~\bibnamefont{Gruber}}, \bibnamefont{and}
  \bibinfo{author}{\bibfnamefont{J.}~\bibnamefont{Wrachtrup}},
  \bibinfo{journal}{Phys. Rev. Lett.} \textbf{\bibinfo{volume}{93}},
  \bibinfo{pages}{130501} (\bibinfo{year}{2004}),
  \urlprefix\url{http://link.aps.org/doi/10.1103/PhysRevLett.93.130501}.

\bibitem[{\citenamefont{Dutt et~al.}(2007)\citenamefont{Dutt, Childress, Jiang,
  Togan, Maze, Jelezko, Zibrov, Hemmer, and Lukin}}]{Dutt2007}
\bibinfo{author}{\bibfnamefont{M.~V.~G.} \bibnamefont{Dutt}},
  \bibinfo{author}{\bibfnamefont{L.}~\bibnamefont{Childress}},
  \bibinfo{author}{\bibfnamefont{L.}~\bibnamefont{Jiang}},
  \bibinfo{author}{\bibfnamefont{E.}~\bibnamefont{Togan}},
  \bibinfo{author}{\bibfnamefont{J.}~\bibnamefont{Maze}},
  \bibinfo{author}{\bibfnamefont{F.}~\bibnamefont{Jelezko}},
  \bibinfo{author}{\bibfnamefont{A.~S.} \bibnamefont{Zibrov}},
  \bibinfo{author}{\bibfnamefont{P.~R.} \bibnamefont{Hemmer}},
  \bibnamefont{and} \bibinfo{author}{\bibfnamefont{M.~D.} \bibnamefont{Lukin}},
  \bibinfo{journal}{Science} \textbf{\bibinfo{volume}{316}},
  \bibinfo{pages}{1312} (\bibinfo{year}{2007}),
  \urlprefix\url{http://www.sciencemag.org/content/316/5829/1312.abstract}.

\bibitem[{\citenamefont{Neumann et~al.}(2008)\citenamefont{Neumann, Mizuochi,
  Rempp, Hemmer, Watanabe, Yamasaki, Jacques, Gaebel, Jelezko, and
  Wrachtrup}}]{Neumann2008}
\bibinfo{author}{\bibfnamefont{P.}~\bibnamefont{Neumann}},
  \bibinfo{author}{\bibfnamefont{N.}~\bibnamefont{Mizuochi}},
  \bibinfo{author}{\bibfnamefont{F.}~\bibnamefont{Rempp}},
  \bibinfo{author}{\bibfnamefont{P.}~\bibnamefont{Hemmer}},
  \bibinfo{author}{\bibfnamefont{H.}~\bibnamefont{Watanabe}},
  \bibinfo{author}{\bibfnamefont{S.}~\bibnamefont{Yamasaki}},
  \bibinfo{author}{\bibfnamefont{V.}~\bibnamefont{Jacques}},
  \bibinfo{author}{\bibfnamefont{T.}~\bibnamefont{Gaebel}},
  \bibinfo{author}{\bibfnamefont{F.}~\bibnamefont{Jelezko}}, \bibnamefont{and}
  \bibinfo{author}{\bibfnamefont{J.}~\bibnamefont{Wrachtrup}},
  \bibinfo{journal}{Science} \textbf{\bibinfo{volume}{320}},
  \bibinfo{pages}{1326} (\bibinfo{year}{2008}),
  \urlprefix\url{http://www.sciencemag.org/content/320/5881/1326.abstract}.

\bibitem[{\citenamefont{Balasubramanian
  et~al.}(2009)\citenamefont{Balasubramanian, Neumann, Twitchen, Markham,
  Kolesov, Mizuochi, Isoya, Achard, Beck, Tissler
  et~al.}}]{Balasubramanian2009}
\bibinfo{author}{\bibfnamefont{G.}~\bibnamefont{Balasubramanian}},
  \bibinfo{author}{\bibfnamefont{P.}~\bibnamefont{Neumann}},
  \bibinfo{author}{\bibfnamefont{D.}~\bibnamefont{Twitchen}},
  \bibinfo{author}{\bibfnamefont{M.}~\bibnamefont{Markham}},
  \bibinfo{author}{\bibfnamefont{R.}~\bibnamefont{Kolesov}},
  \bibinfo{author}{\bibfnamefont{N.}~\bibnamefont{Mizuochi}},
  \bibinfo{author}{\bibfnamefont{J.}~\bibnamefont{Isoya}},
  \bibinfo{author}{\bibfnamefont{J.}~\bibnamefont{Achard}},
  \bibinfo{author}{\bibfnamefont{J.}~\bibnamefont{Beck}},
  \bibinfo{author}{\bibfnamefont{J.}~\bibnamefont{Tissler}},
  \bibnamefont{et~al.}, \bibinfo{journal}{Nat Mater}
  \textbf{\bibinfo{volume}{8}}, \bibinfo{pages}{383} (\bibinfo{year}{2009}),
  \urlprefix\url{http://dx.doi.org/10.1038/nmat2420}.

\bibitem[{\citenamefont{Maze et~al.}(2008{\natexlab{a}})\citenamefont{Maze,
  Stanwix, Hodges, Hong, Taylor, Cappellaro, Jiang, Dutt, Togan, Zibrov
  et~al.}}]{Maze2008a}
\bibinfo{author}{\bibfnamefont{J.~R.} \bibnamefont{Maze}},
  \bibinfo{author}{\bibfnamefont{P.~L.} \bibnamefont{Stanwix}},
  \bibinfo{author}{\bibfnamefont{J.~S.} \bibnamefont{Hodges}},
  \bibinfo{author}{\bibfnamefont{S.}~\bibnamefont{Hong}},
  \bibinfo{author}{\bibfnamefont{J.~M.} \bibnamefont{Taylor}},
  \bibinfo{author}{\bibfnamefont{P.}~\bibnamefont{Cappellaro}},
  \bibinfo{author}{\bibfnamefont{L.}~\bibnamefont{Jiang}},
  \bibinfo{author}{\bibfnamefont{M.~V.~G.} \bibnamefont{Dutt}},
  \bibinfo{author}{\bibfnamefont{E.}~\bibnamefont{Togan}},
  \bibinfo{author}{\bibfnamefont{A.~S.} \bibnamefont{Zibrov}},
  \bibnamefont{et~al.}, \bibinfo{journal}{Nature}
  \textbf{\bibinfo{volume}{455}}, \bibinfo{pages}{644}
  (\bibinfo{year}{2008}{\natexlab{a}}),
  \urlprefix\url{http://dx.doi.org/10.1038/nature07279}.

\bibitem[{\citenamefont{Dolde et~al.}(2011)\citenamefont{Dolde, Fedder,
  Doherty, Nobauer, Rempp, Balasubramanian, Wolf, Reinhard, Hollenberg, Jelezko
  et~al.}}]{Dolde2011}
\bibinfo{author}{\bibfnamefont{F.}~\bibnamefont{Dolde}},
  \bibinfo{author}{\bibfnamefont{H.}~\bibnamefont{Fedder}},
  \bibinfo{author}{\bibfnamefont{M.~W.} \bibnamefont{Doherty}},
  \bibinfo{author}{\bibfnamefont{T.}~\bibnamefont{Nobauer}},
  \bibinfo{author}{\bibfnamefont{F.}~\bibnamefont{Rempp}},
  \bibinfo{author}{\bibfnamefont{G.}~\bibnamefont{Balasubramanian}},
  \bibinfo{author}{\bibfnamefont{T.}~\bibnamefont{Wolf}},
  \bibinfo{author}{\bibfnamefont{F.}~\bibnamefont{Reinhard}},
  \bibinfo{author}{\bibfnamefont{L.~C.~L.} \bibnamefont{Hollenberg}},
  \bibinfo{author}{\bibfnamefont{F.}~\bibnamefont{Jelezko}},
  \bibnamefont{et~al.}, \bibinfo{journal}{Nat Phys}
  \textbf{\bibinfo{volume}{7}}, \bibinfo{pages}{459} (\bibinfo{year}{2011}),
  \urlprefix\url{http://dx.doi.org/10.1038/nphys1969}.

\bibitem[{\citenamefont{Barrett and Kok}(2005)}]{Barrett2005}
\bibinfo{author}{\bibfnamefont{S.~D.} \bibnamefont{Barrett}} \bibnamefont{and}
  \bibinfo{author}{\bibfnamefont{P.}~\bibnamefont{Kok}},
  \bibinfo{journal}{Phys. Rev. A} \textbf{\bibinfo{volume}{71}},
  \bibinfo{pages}{060310} (\bibinfo{year}{2005}),
  \urlprefix\url{http://link.aps.org/doi/10.1103/PhysRevA.71.060310}.

\bibitem[{\citenamefont{Stoneham et~al.}(2009)\citenamefont{Stoneham, Harker,
  and Morley}}]{Stoneham2009}
\bibinfo{author}{\bibfnamefont{A.~M.} \bibnamefont{Stoneham}},
  \bibinfo{author}{\bibfnamefont{A.~H.} \bibnamefont{Harker}},
  \bibnamefont{and} \bibinfo{author}{\bibfnamefont{G.~W.}
  \bibnamefont{Morley}}, \bibinfo{journal}{Journal of Physics: Condensed
  Matter} \textbf{\bibinfo{volume}{21}}, \bibinfo{pages}{364222}
  (\bibinfo{year}{2009}),
  \urlprefix\url{http://stacks.iop.org/0953-8984/21/i=36/a=364222}.

\bibitem[{\citenamefont{Smeltzer et~al.}(2011)\citenamefont{Smeltzer,
  Childress, and Gali}}]{Smeltzer2011}
\bibinfo{author}{\bibfnamefont{B.}~\bibnamefont{Smeltzer}},
  \bibinfo{author}{\bibfnamefont{L.}~\bibnamefont{Childress}},
  \bibnamefont{and} \bibinfo{author}{\bibfnamefont{A.}~\bibnamefont{Gali}},
  \bibinfo{journal}{New Journal of Physics} \textbf{\bibinfo{volume}{13}},
  \bibinfo{pages}{025021} (\bibinfo{year}{2011}),
  \urlprefix\url{http://stacks.iop.org/1367-2630/13/i=2/a=025021}.

\bibitem[{\citenamefont{Morley et~al.}(2012)\citenamefont{Morley, Lueders,
  Hamed~Mohammady, Balian, Aeppli, Kay, Witzel, Jeschke, and
  Monteiro}}]{Morley2012}
\bibinfo{author}{\bibfnamefont{G.~W.} \bibnamefont{Morley}},
  \bibinfo{author}{\bibfnamefont{P.}~\bibnamefont{Lueders}},
  \bibinfo{author}{\bibfnamefont{M.}~\bibnamefont{Hamed~Mohammady}},
  \bibinfo{author}{\bibfnamefont{S.~J.} \bibnamefont{Balian}},
  \bibinfo{author}{\bibfnamefont{G.}~\bibnamefont{Aeppli}},
  \bibinfo{author}{\bibfnamefont{C.~W.~M.} \bibnamefont{Kay}},
  \bibinfo{author}{\bibfnamefont{W.~M.} \bibnamefont{Witzel}},
  \bibinfo{author}{\bibfnamefont{G.}~\bibnamefont{Jeschke}}, \bibnamefont{and}
  \bibinfo{author}{\bibfnamefont{T.~S.} \bibnamefont{Monteiro}},
  \bibinfo{journal}{Nat Mater} \textbf{\bibinfo{volume}{advance online
  publication}},  (\bibinfo{year}{2012}), ISSN \bibinfo{issn}{1476-4660},
  \urlprefix\url{http://dx.doi.org/10.1038/nmat3499}.

\bibitem[{\citenamefont{Bertaina et~al.}(2009)\citenamefont{Bertaina, Shim,
  Gambarelli, Malkin, and Barbara}}]{Bertaina2009}
\bibinfo{author}{\bibfnamefont{S.}~\bibnamefont{Bertaina}},
  \bibinfo{author}{\bibfnamefont{J.~H.} \bibnamefont{Shim}},
  \bibinfo{author}{\bibfnamefont{S.}~\bibnamefont{Gambarelli}},
  \bibinfo{author}{\bibfnamefont{B.~Z.} \bibnamefont{Malkin}},
  \bibnamefont{and} \bibinfo{author}{\bibfnamefont{B.}~\bibnamefont{Barbara}},
  \bibinfo{journal}{Phys. Rev. Lett.} \textbf{\bibinfo{volume}{103}},
  \bibinfo{pages}{226402} (\bibinfo{year}{2009}),
  \urlprefix\url{http://link.aps.org/doi/10.1103/PhysRevLett.103.226402}.

\bibitem[{\citenamefont{Turov and Petrov}(1972)}]{Turov}
\bibinfo{author}{\bibfnamefont{E.}~\bibnamefont{Turov}} \bibnamefont{and}
  \bibinfo{author}{\bibfnamefont{M.~P.} \bibnamefont{Petrov}},
  \emph{\bibinfo{title}{Nuclear magnetic resonance in ferro-and
  antiferromagnets}} (\bibinfo{publisher}{Israel Program for Scientific
  Translations}, \bibinfo{year}{1972}).

\bibitem[{\citenamefont{Maze et~al.}(2008{\natexlab{b}})\citenamefont{Maze,
  Taylor, and Lukin}}]{Maze2008}
\bibinfo{author}{\bibfnamefont{J.~R.} \bibnamefont{Maze}},
  \bibinfo{author}{\bibfnamefont{J.~M.} \bibnamefont{Taylor}},
  \bibnamefont{and} \bibinfo{author}{\bibfnamefont{M.~D.} \bibnamefont{Lukin}},
  \bibinfo{journal}{Phys. Rev. B} \textbf{\bibinfo{volume}{78}},
  \bibinfo{pages}{094303} (\bibinfo{year}{2008}{\natexlab{b}}),
  \urlprefix\url{http://link.aps.org/doi/10.1103/PhysRevB.78.094303}.

\bibitem[{Shi()}]{Shim_Prepare}
\bibinfo{note}{A paper about the precise measurement of $A_{\parallel}$ and
  $A_{\perp}$ is in preparation.}

\bibitem[{\citenamefont{Jacques et~al.}(2009)\citenamefont{Jacques, Neumann,
  Beck, Markham, Twitchen, Meijer, Kaiser, Balasubramanian, Jelezko, and
  Wrachtrup}}]{Jacques2009}
\bibinfo{author}{\bibfnamefont{V.}~\bibnamefont{Jacques}},
  \bibinfo{author}{\bibfnamefont{P.}~\bibnamefont{Neumann}},
  \bibinfo{author}{\bibfnamefont{J.}~\bibnamefont{Beck}},
  \bibinfo{author}{\bibfnamefont{M.}~\bibnamefont{Markham}},
  \bibinfo{author}{\bibfnamefont{D.}~\bibnamefont{Twitchen}},
  \bibinfo{author}{\bibfnamefont{J.}~\bibnamefont{Meijer}},
  \bibinfo{author}{\bibfnamefont{F.}~\bibnamefont{Kaiser}},
  \bibinfo{author}{\bibfnamefont{G.}~\bibnamefont{Balasubramanian}},
  \bibinfo{author}{\bibfnamefont{F.}~\bibnamefont{Jelezko}}, \bibnamefont{and}
  \bibinfo{author}{\bibfnamefont{J.}~\bibnamefont{Wrachtrup}},
  \bibinfo{journal}{Phys. Rev. Lett.} \textbf{\bibinfo{volume}{102}},
  \bibinfo{pages}{057403} (\bibinfo{year}{2009}),
  \urlprefix\url{http://link.aps.org/doi/10.1103/PhysRevLett.102.057403}.

\bibitem[{\citenamefont{Neumann et~al.}(2010)\citenamefont{Neumann, Beck,
  Steiner, Rempp, Fedder, Hemmer, Wrachtrup, and Jelezko}}]{Neumann2010}
\bibinfo{author}{\bibfnamefont{P.}~\bibnamefont{Neumann}},
  \bibinfo{author}{\bibfnamefont{J.}~\bibnamefont{Beck}},
  \bibinfo{author}{\bibfnamefont{M.}~\bibnamefont{Steiner}},
  \bibinfo{author}{\bibfnamefont{F.}~\bibnamefont{Rempp}},
  \bibinfo{author}{\bibfnamefont{H.}~\bibnamefont{Fedder}},
  \bibinfo{author}{\bibfnamefont{P.~R.} \bibnamefont{Hemmer}},
  \bibinfo{author}{\bibfnamefont{J.}~\bibnamefont{Wrachtrup}},
  \bibnamefont{and} \bibinfo{author}{\bibfnamefont{F.}~\bibnamefont{Jelezko}},
  \bibinfo{journal}{Science} \textbf{\bibinfo{volume}{329}},
  \bibinfo{pages}{542} (\bibinfo{year}{2010}),
  \urlprefix\url{http://www.sciencemag.org/content/329/5991/542.abstract}.

\bibitem[{\citenamefont{Robledo et~al.}(2011)\citenamefont{Robledo, Childress,
  Bernien, Hensen, Alkemade, and Hanson}}]{Robledo2011}
\bibinfo{author}{\bibfnamefont{L.}~\bibnamefont{Robledo}},
  \bibinfo{author}{\bibfnamefont{L.}~\bibnamefont{Childress}},
  \bibinfo{author}{\bibfnamefont{H.}~\bibnamefont{Bernien}},
  \bibinfo{author}{\bibfnamefont{B.}~\bibnamefont{Hensen}},
  \bibinfo{author}{\bibfnamefont{P.~F.~A.} \bibnamefont{Alkemade}},
  \bibnamefont{and} \bibinfo{author}{\bibfnamefont{R.}~\bibnamefont{Hanson}},
  \bibinfo{journal}{Nature} \textbf{\bibinfo{volume}{477}},
  \bibinfo{pages}{574} (\bibinfo{year}{2011}),
  \urlprefix\url{http://dx.doi.org/10.1038/nature10401}.

\bibitem[{\citenamefont{Jiang et~al.}(2009)\citenamefont{Jiang, Hodges, Maze,
  Maurer, Taylor, Cory, Hemmer, Walsworth, Yacoby, Zibrov et~al.}}]{Jiang2009}
\bibinfo{author}{\bibfnamefont{L.}~\bibnamefont{Jiang}},
  \bibinfo{author}{\bibfnamefont{J.~S.} \bibnamefont{Hodges}},
  \bibinfo{author}{\bibfnamefont{J.~R.} \bibnamefont{Maze}},
  \bibinfo{author}{\bibfnamefont{P.}~\bibnamefont{Maurer}},
  \bibinfo{author}{\bibfnamefont{J.~M.} \bibnamefont{Taylor}},
  \bibinfo{author}{\bibfnamefont{D.~G.} \bibnamefont{Cory}},
  \bibinfo{author}{\bibfnamefont{P.~R.} \bibnamefont{Hemmer}},
  \bibinfo{author}{\bibfnamefont{R.~L.} \bibnamefont{Walsworth}},
  \bibinfo{author}{\bibfnamefont{A.}~\bibnamefont{Yacoby}},
  \bibinfo{author}{\bibfnamefont{A.~S.} \bibnamefont{Zibrov}},
  \bibnamefont{et~al.}, \bibinfo{journal}{Science}
  \textbf{\bibinfo{volume}{326}}, \bibinfo{pages}{267} (\bibinfo{year}{2009}),
  \urlprefix\url{http://www.sciencemag.org/content/326/5950/267.abstract}.

\bibitem[{Note1()}]{Note1}
Note1, \bibinfo{note}{The initial part of the FID is distorted by baseline
  drifts and compressed amplitudes}.

\bibitem[{\citenamefont{Ryan et~al.}(2010)\citenamefont{Ryan, Hodges, and
  Cory}}]{Ryan2010}
\bibinfo{author}{\bibfnamefont{C.~A.} \bibnamefont{Ryan}},
  \bibinfo{author}{\bibfnamefont{J.~S.} \bibnamefont{Hodges}},
  \bibnamefont{and} \bibinfo{author}{\bibfnamefont{D.~G.} \bibnamefont{Cory}},
  \bibinfo{journal}{Phys. Rev. Lett.} \textbf{\bibinfo{volume}{105}},
  \bibinfo{pages}{200402} (\bibinfo{year}{2010}),
  \urlprefix\url{http://link.aps.org/doi/10.1103/PhysRevLett.105.200402}.

\bibitem[{\citenamefont{Naydenov et~al.}(2011)\citenamefont{Naydenov, Dolde,
  Hall, Shin, Fedder, Hollenberg, Jelezko, and Wrachtrup}}]{Naydenov2011}
\bibinfo{author}{\bibfnamefont{B.}~\bibnamefont{Naydenov}},
  \bibinfo{author}{\bibfnamefont{F.}~\bibnamefont{Dolde}},
  \bibinfo{author}{\bibfnamefont{L.~T.} \bibnamefont{Hall}},
  \bibinfo{author}{\bibfnamefont{C.}~\bibnamefont{Shin}},
  \bibinfo{author}{\bibfnamefont{H.}~\bibnamefont{Fedder}},
  \bibinfo{author}{\bibfnamefont{L.~C.~L.} \bibnamefont{Hollenberg}},
  \bibinfo{author}{\bibfnamefont{F.}~\bibnamefont{Jelezko}}, \bibnamefont{and}
  \bibinfo{author}{\bibfnamefont{J.}~\bibnamefont{Wrachtrup}},
  \bibinfo{journal}{Phys. Rev. B} \textbf{\bibinfo{volume}{83}},
  \bibinfo{pages}{081201} (\bibinfo{year}{2011}),
  \urlprefix\url{http://link.aps.org/doi/10.1103/PhysRevB.83.081201}.

\bibitem[{\citenamefont{Bar-Gill et~al.}(2012)\citenamefont{Bar-Gill, Pham,
  Belthangady, Le~Sage, Cappellaro, Maze, Lukin, Yacoby, and
  Walsworth}}]{Bar-Gill2012}
\bibinfo{author}{\bibfnamefont{N.}~\bibnamefont{Bar-Gill}},
  \bibinfo{author}{\bibfnamefont{L.}~\bibnamefont{Pham}},
  \bibinfo{author}{\bibfnamefont{C.}~\bibnamefont{Belthangady}},
  \bibinfo{author}{\bibfnamefont{D.}~\bibnamefont{Le~Sage}},
  \bibinfo{author}{\bibfnamefont{P.}~\bibnamefont{Cappellaro}},
  \bibinfo{author}{\bibfnamefont{J.}~\bibnamefont{Maze}},
  \bibinfo{author}{\bibfnamefont{M.}~\bibnamefont{Lukin}},
  \bibinfo{author}{\bibfnamefont{A.}~\bibnamefont{Yacoby}}, \bibnamefont{and}
  \bibinfo{author}{\bibfnamefont{R.}~\bibnamefont{Walsworth}},
  \bibinfo{journal}{Nat Commun} \textbf{\bibinfo{volume}{3}},
  \bibinfo{pages}{858} (\bibinfo{year}{2012}),
  \urlprefix\url{http://dx.doi.org/10.1038/ncomms1856}.

\bibitem[{\citenamefont{Shim et~al.}(2012)\citenamefont{Shim, Niemeyer, Zhang,
  and Suter}}]{Shim2012}
\bibinfo{author}{\bibfnamefont{J.~H.} \bibnamefont{Shim}},
  \bibinfo{author}{\bibfnamefont{I.}~\bibnamefont{Niemeyer}},
  \bibinfo{author}{\bibfnamefont{J.}~\bibnamefont{Zhang}}, \bibnamefont{and}
  \bibinfo{author}{\bibfnamefont{D.}~\bibnamefont{Suter}},
  \bibinfo{journal}{Europhys. Lett.} \textbf{\bibinfo{volume}{99}},
  \bibinfo{pages}{40004} (\bibinfo{year}{2012}),
  \urlprefix\url{http://stacks.iop.org/0295-5075/99/i=4/a=40004}.

\end{thebibliography}

\end{document}